\documentclass[a4paper,12pt]{article}

\usepackage[center]{titlesec}

\renewcommand{\thesection}{\Roman{section}}

\renewcommand{\thesubsection}{\arabic{subsection}}

\renewcommand{\thesubsubsection}{\alph{subsubsection}}

\titleformat%
    {\section}
    [block]
    {\centering\bfseries}
    {\thesection.}
    {0.5em}
    {}

\titleformat%
    {\subsection}
    [block]
    {\centering\itshape}
    {\thesubsection.}
    {0.5em}
    {}

\titleformat%
    {\subsubsection}
    [block]
    {\centering}
    {\thesubsubsection.)}
    {0.5em}
    {}

\usepackage{lipsum}

\titlespacing{\section}{0pt}{*4}{*1.5}
\titlespacing{\subsection}{0pt}{*4}{*1.5}
\titlespacing{\subsubsection}{0pt}{*4}{*1.5}

\usepackage[utf8]{inputenc}
\usepackage{cancel}
\usepackage{tikz}
\usepackage{ulem}
\usepackage{amsfonts}
\usepackage{amssymb}
\usepackage{graphicx}
\usepackage{amsmath}
\usepackage{enumerate}
\usepackage{mathtools}
\usepackage{subfig}
\usepackage{color}
\usepackage{tikz}
\usepackage{float}
\usepackage{here}
\usepackage{cite}
\usepackage{mathrsfs}
\usepackage{float,epsfig}
\usepackage{dcolumn}
\usepackage{graphicx}
\usepackage{bm}
\usepackage{amsmath,amssymb,amsthm}
\usepackage[colorlinks=true,linkcolor=blue,citecolor=red]{hyperref}
\usepackage{multirow}
\usepackage[toc,page]{appendix}
\usetikzlibrary{arrows.meta}
\usetikzlibrary{bending}
\usetikzlibrary{calc}
\newcommand{\be}{\begin{equation}}
\newcommand{\ee}{\end{equation}}
\newcommand{\bea}{\setlength\arraycolsep{2pt} \begin{eqnarray}}
\newcommand{\eea}{\end{eqnarray}}

\def\0{{\sst{(0)}}}
\def\1{{\sst{(1)}}}
\def\2{{\sst{(2)}}}
\def\3{{\sst{(3)}}}
\def\4{{\sst{(4)}}}
\def\5{{\sst{(5)}}}
\def\6{{\sst{(6)}}}
\def\7{{\sst{(7)}}}
\def\8{{\sst{(8)}}}
\def\sst#1{{\scriptscriptstyle #1}}

\makeatletter \@addtoreset{equation}{section}

\definecolor{lime}{HTML}{A6CE39}

\begin{document}

\title{{\normalsize \textbf{\Large   CUDA Assisted Swampland and  Black Hole  Thermodynamics     }}}
\author{ {\small  Saad Eddine Baddis\footnote{Corresponding author: saadeddine.baddis@um5r.ac.ma},     Adil  Belhaj,    Hajar Belmahi\thanks{
 Authors are  in alphabetical order.} \hspace*{-8pt}} \\
{\small  ESMaR, Faculty of Science, Mohammed V University in Rabat, Rabat, Morocco}}
\maketitle

\begin{abstract}
Motivated by string theory activities, we investigate  the swampland program in the  black hole  context  via CUDA numerical computations.  Precisely,   we study   charged black hole solutions submerged in a hypergeometric inflationary  potential extracted from  Gauss-Bonnet scalar couplings  to the Einstein–Maxwell–Hilbert action.  Exploiting CUDA enabled parallel programming techniques,  we examine the scalar  potential  behaviors permitting  to derive the physically relevant roots of the black hole  metric function. Equipped with more powerful CUDA techniques, we  explore  the effects  of the  parametric quantities  on such roots supporting a swampland investigation.  Accordingly,  we establish a relationship between the charge $Q$ and  the mass $M$  of a  charged black hole allowing   to approach  the extremal   limit and the cosmic horizon behaviors.  Furthermore, we highlight the implications of the scalar field related to swampland conjectures including the  moduli distance. Employing   certain  developed CUDA techniques to extract criticality conditions via black hole thermodynamics,  we establish a relationship between the scalar moduli  and the charge $Q$, enabling  to show at what stage black holes can become unstable disintegrating  into light states associated with the ratio $\frac{Q}{M}>2\sqrt{\pi}$.

\end{abstract}

\newpage

\section{Introduction}

Recently,  a special emphasis has  been put on the   swampland  program. The latter  is an initiative aimed at identifying  what an effective field theory (EFT) compatible with quantum gravity (QG)  should look like. Through several assumptions, this program explores aspects of falsification scenario in the EFT in question. More specifically, the falsifiability of an EFT is assessed by recognizing   the regions incompatible with quantum gravity theory (QGT)  in the corresponding moduli space (the “swampland”) and the regions compatible with  QGT  (the “landscape”). More specifically, these criteria provide a qualitative idea of EFT aspects \cite{SL1,SL2,SL3,SL4,SL5,SL6}.   In this way, various  conjectures have been proposed and elaborated in connection with certain physical theories including superstring models. One of them  is the distance conjecture, which prevents a  low-energy EFT from accessing infinite towers of light states, thereby effectively eliminating the  decompactification scenarios.  In addition, this conjecture has been investigated in the context of T-duality and its extended version called  mirror symmetry in the case of Calabi-Yau manifold  compactifications  producing various scalar fields in lower dimensional stringy spectrums \cite{DC1,DC2,DC3,DC4,DC5,DC6,DC7,DC8,DC9,DC10}.  In concrete terms, this   can be exploited to approach   the string landscape problem.  Such fields determine  the stability of a de Sitter space by studying the de Sitter conjecture \cite{dSC1,dSC2,dSC3}. Going further, the fact that the observable universe is in a vacuum translates into the absence of run away instabilities. Similarly, tachyonic instabilities are eliminated using the associated Hessian condition.
Indeed, with the exception of a few special cases, this conjecture has been found to be inconsistent with inflationary scenarios \cite{SL2}. However, certain inflationary models developed in \cite{dSC4,dSC6} and hypergeometric inflationary potentials in the context of string corrections \cite{dSC7} have been found to be compatible without any particular adjustments.  Alternatively, the polynomial potential obtained from  such stringy  corrections has been observed to be in contradiction with the swampland \cite{dSC8}.\\
Another feature of EFT compatibility with  QGT is that the restoration of global gauge symmetries is prohibited by a scale at which the low-energy EFT breaks down. This new scale appears to be limited from above by the gauge coupling. In addition, the removal of black hole remnants is a concomitant feature \cite{WGC1,WGC2,WGC3,WGC5,WGC6,WGC7,WGC8,WGC9}. More specifically, the  black holes eventually disintegrate into light particles with maximum charges, due to instabilities at their extrema limit.  Effectively, this could  suppress the undesirable behaviors associated with  the naked singularities \cite{Rem1,Rem2}.\\ These two considerations are grouped together in what is known as the weak gravity conjecture (WGC).  As its name suggests, this conjecture is based on the fact that gravity is the weakest force, which could explain  the  new scale limit. One  fundamental region that could be interesting  to  validate  this hypothesis  concerns  event horizons, where  the accessible gravitational effects are at their extreme.  A verification of this kind has been carried out in M-theory using the recombination factor in  different  Calabi-Yau threefold  compactification constructions with lower  dimensional Kahler  moduli spaces \cite{WGCCY1}, where the instabilities associated with WGC have been described for BPS and non-BPS black holes and black strings.

The study of the event horizon of certain black holes  coupled to gauge models  in an inflationary dS space  may require intensive numerical computations. These include algorithms executed in parallel using
CUDA. The latter is considered as a  general purpose parallel computing platform and programming model that leverages the parallel compute engine in NVIDIA GPUs \cite{CUDA1,CUDA2,CUDA3,CUDA4}. In addition, these GPUs introduce a multitude of powerful  methods and efficient tools to further exploit the associated  subtleties. Naturally, this type of numerical analysis has been widely used in black hole physics. Particularly, this includes  the study of black hole shadows and other cosmic phenomena \cite{CUDABH1,CUDABH2,CUDABH3}.\\

The main  objective of this work   is to implement   CUDA numerical computations in   the swampland program via   black hole physics. Concretely,   we   investigate the    charged black hole solutions submerged in a hypergeometric inflationary  potential  obtained  from  Gauss-Bonnet scalar couplings  to the Maxwell-Einstein-Hilbert action.  Exploiting CUDA enabled parallel programming techniques,  we  reconsider the study of   the scalar  potential  behaviors allowing to  provide  certain   relevant roots of the black hole  metric function. Equipped with more powerful CUDA techniques, we  explore the effects of the  parametric quantities  on such roots supporting a swampland program  investigation.  Accordingly,  we  provide  a relationship between the electric  charge $Q$ and  the mass $M$  of a    stringy  charged  black hole permitting    to approach the  extremal  limit and the cosmic horizon behaviors.  Furthermore, we unveil the scalar field implications making contact with swampland conjectures regarding the moduli distance. Exploiting the developed CUDA techniques to extract criticality conditions via the  black hole thermodynamics, we establish a close relationship between the scalar  moduli and the   charge enabling to  show at which stage black holes can become unstable and decay to light states of the ratio $\frac{Q}{M}>2\sqrt{\pi}$.

The organization of this work is as follows. In section 2, we  reconsider the study of charged black holes in the  Gauss-Bonnet coupling  scenarios.   Section 3 concerns  the investigation the charge to the mass ration in the swampland  conjecture context.  In section 4,   we   bridge the swampland program with  the black hole    thermodynamics  via CUDA  computational techniques. The last section is devoted to concluding remarks and certain open questions. 

\section{On stringy  black holes }
\label{sec:2}
\begin{figure}
\begin{center}
\includegraphics[scale=0.7]{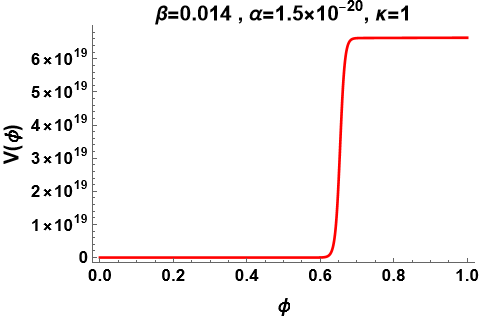}
\caption{Hypergeometric scalar potential for $m=2$, and $n=1$}
\label{Fig1}
\end{center}
\end{figure}

In this section, we reconsider the study of charged black holes in the inflationary theory scenarios. Specifically,  such inflation  models rely on  hypergeometric scalar potentials extracted from stringy corrected gravity theories.  These gravity theories  are usually described by 
  minimal Maxwell-Einstein-Hilbert  contributions  with  corrections motivated  by string theory and related topics including M-theory.  They  involve couplings  
of  scalar fields  to the Einstein tensor   and the   4-dimensional  Gauss-Bonnet (GB)  term via   differentiable functions. The introduction of these coupling functions has been  presented   in many places  where they have been considered as contributions of high order curvature terms, also known as corrections to the Maxwell-Einstein-Hilbert action   \cite{3,2}. Taking  $M_p^{-2}=8\pi G=1$,   the corresponding   action, in the Einstein frame,  can be  expressed as follows
\begin{equation}
\label{e0021}
S=\int d^4x\sqrt{-g}\left[\frac{R }{2\kappa^2}-\frac{1}{4}F^{\mu\nu}F_{\mu\nu}-\frac{1}{2}\partial_{\mu}\phi\partial^{\mu}\phi-V(\phi)-\xi(\phi){\cal G}\right], 
\end{equation}
where $F^{\mu\nu}F_{\mu\nu}$ is the Maxwell tensor,  and $\phi$ is a real scalar field with a potential $V(\phi)$.   ${\cal G}$  represents  the GB 4-dimensional invariant  term given by
${\cal G}=R^2-4R_{\mu\nu}R^{\mu\nu}+R_{\mu\nu\lambda\rho}R^{\mu\nu\lambda\rho}$,
 where  $R$ is the Ricci scalar. 
 $\xi$ is a real   differentiable function of $\phi$   describing the stringy corrections to the ordinary  gravity.    The above action provides a  family of models  
specified by   two   scalar   functions  $V$ and $\xi$.  A priori these functions  should be arbitrary.  However, here,  we consider functions providing  cosmological behaviors. Indeed,  the conservation condition on the action Eq.(\ref{e0021}) gives the following Einstein field equations
\begin{equation}
R_{\mu\nu}-(g_{\mu\nu}\nabla^\sigma\nabla_\sigma-\nabla_{\mu}\nabla_{\nu})\xi(\phi)-\frac{1}{2}g_{\mu\nu}R=\kappa^2T_{\mu\nu},
\end{equation}
where the tensor $T_{\mu\nu}$ containing matter contributions  is given  by 
\begin{equation}
T_{\mu\nu} = \partial_\mu\phi\partial_\nu\phi-\frac{1}{2}g_{\mu\nu}\partial_\sigma\phi\partial^\sigma\phi+F_{\mu\sigma}F^{\sigma}_{\mu}-\frac{1}{4}g_{\mu\nu}F_{\rho\sigma}F^{\rho\sigma}-g_{\mu\nu}\frac{V(\phi)}{k^2}.
\end{equation}
 To elaborate  corroborated models 
 matching with   the observational  data  \cite{4,5,6,7,8,11}, certain requirements should be imposed on the   scalar quantities.  Indeed, we consider the following condition on the  dynamical scalar field
\begin{eqnarray}
\phi &\equiv & \phi(t)\\
\
\partial_t\phi\partial^t\phi &\approx &  0
\label{e0022}
\end{eqnarray}
being  compatible with the slow-roll conditions
\begin{equation}
\dot{\phi}^2<<(V(\phi)\leq\mathcal{O}(1)).
\end{equation}
It is denoted in passing that generic scalar field configurations  could be considered  supplying discussions going beyond the scope of the present work.  Certain details concerning  extended configurations  could be found elsewhere.
Considering the previous scalar field constraints,   the assumed  ansatz  of the  black hole line element can    take the following  form 
\begin{eqnarray}
ds^2 = g(r)dt^2-\frac{1}{g(r)}dr^2-r^2d\Omega^2.
\end{eqnarray} 
The condition Eq.(\ref{e0022})  provides an exact form of the black hole metric function 
\begin{eqnarray}
g(r) &=& 1-\frac{2M}{r}+\frac{Q^2}{4\pi r^2}-V(\phi)r^2
\label{Eq}
\end{eqnarray} 
where one has used  the normal unites $G=c=\epsilon_0=\kappa=1$. It has been remarked that the underlying physical discussion depends on the  scalar potential form.   It has been shown that  such  forms can be derived  from  the scalar coupling function using  some techniques developed in  \cite{dSC7}.  A priori, there are many possibilities. However,    we restrict ourselves  to  the scalar coupling function  expressed as  follows 
\begin{eqnarray}
\xi(\phi) = \lambda\int^{\phi}_0 e^{-\frac{1}{\beta}(\frac{m}{m+n}x^{m+n}+x^m)}dx
\end{eqnarray}
where $\lambda$  is a normalized  parameter and $\beta$ is a free parameter.  Exploiting such techniques,  this generates  a  hypergeometric scalar potential given by 
\begin{eqnarray}
\small{V(\phi) = \frac{e^{-\beta\frac{(\phi)^{2-(m+n)}}{\alpha m(m+n-2)} \; _2F_1(1,\frac{m+n-2}{n};\frac{m+2n-2}{n};-(\phi)^{-n})}}{\alpha+\frac{8}{3}\int_{\phi}^{+ \infty}\xi^\prime(x) e^{-\beta\frac{( x)^{2-(m+n)}}{m(m+n-2)} \;_2F_1(1,\frac{m+n-2}{n};\frac{m+2n-2}{n};-( x)^{-n})}dx}}
\end{eqnarray}
where $\alpha$ is an integration parameter.  A close examination  shows  that  the  handling of  such  a scalar potential  may need intensive numerical calculations.  To overcome such  issues, we  deal with the relevant integrals as a memory reduction problem using parallel programming techniques such as warps, shared memory and thread synchronization \cite{CUDA1,CUDA2}. To do so, we  exploit  these techniques to  elaborate  a CUDA code capitalizing  on the GPU architecture. It is denoted in passing that  the CUDA enabled parallel programming methods  have been exploited in many physical problems such as solving complex differential equations, and simulating intricate physical phenomena.  To approach the proposed scalar potential  via CUDA code, we consider a special  case associated with $m=2$ and $n=1$. Using the CUDA obtained data,   Fig.(\ref{Fig1}) illustrates  the scalar potential  for $m=2$ and $n=1$.  
From  this figure,  it has been observed  the slow-roll and the inflationary characteristics.

\begin{figure}
\begin{center}
\includegraphics[scale=0.7]{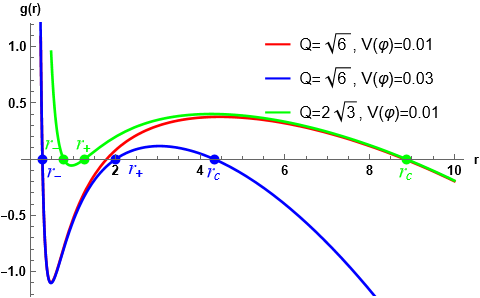}
\caption{Metric function in terms of the radial variable. }
\label{Fig2}
\end{center}
\end{figure}

To extract certain black  hole physical behaviors, the metric function  $g(r)$ should be examined.  A close inspection shows that this function involves four roots. However, one of them is negative  which should be omitted by physical reasons.   The remaining  ones are   positive being expressed as follows
\begin{eqnarray}
r_+ &=& \frac{\sqrt{4+\mathcal{L}-\frac{2M\sqrt{3V(\phi)}}{\sqrt{2-\mathcal{L}}}}+\sqrt{2-\mathcal{L}}}{2\sqrt{3}V(\phi)},\\
r_- &=& \frac{\sqrt{4+\mathcal{L}-\frac{2M\sqrt{3V(\phi)}}{\sqrt{2-\mathcal{L}}}}-\sqrt{2-\mathcal{L}}}{2\sqrt{3}V(\phi)},\\
r_c &=& \frac{\sqrt{4+\mathcal{L}+\frac{2M\sqrt{3V(\phi)}}{\sqrt{2-\mathcal{L}}}}+\sqrt{2-\mathcal{L}}}{2\sqrt{3}V(\phi)},
\end{eqnarray}
where  one has used
\begin{eqnarray}
\mathcal{L} &=& \mathcal{P}(\frac{2}{\psi})^{\frac{1}{3}}+(\frac{\psi}{2})^{\frac{1}{3}},\\
\psi &=& \rho + \sqrt{\rho^2-4\mathcal{P}^3},\\
\rho &=& 2 - V(\phi)(27M^2-18\frac{Q^2}{\pi}),\\
\mathcal{P} &=& 1-3V(\phi)\frac{Q^2}{\pi}.\\
\end{eqnarray}
For certain black hole  parameters,   such solutions are  depicted in  Fig.(\ref{Fig2}) showing the effects of the charge and the scalar field values on the obtained  horizon radii.

 Fixing the scalar potential,   the root $r_+$  decreases  by increasing the charge $Q$  while the opposite behavior is observed for $r_-$. In the other hand, by changing the values of the scalar field,  we  note  that   the root $r_c$ is sent further in  the radial direction while the roots $r_+$ and $r_-$ roughly keep the   same behaviors.  It has been remarked that the cosmic horizon radius   $r_c$ is more susceptible to the scalar field values. However,  the inner and the  outer horizons   are more sensitive to  the electric charge variation.  

\begin{figure}
\begin{center}
\begin{tabbing}
\includegraphics[scale=0.55]{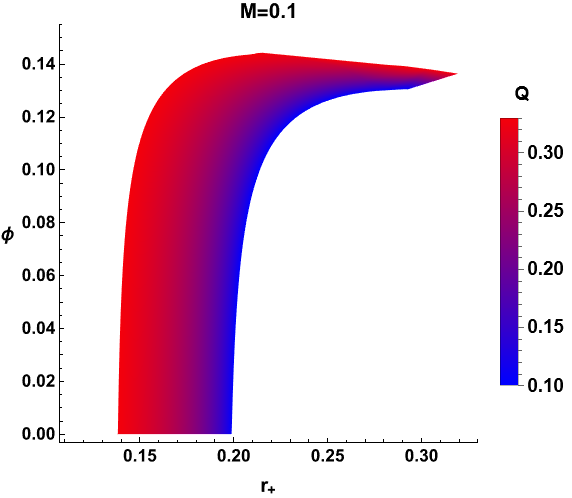}
\hspace{0.2cm} \includegraphics[scale=0.55]{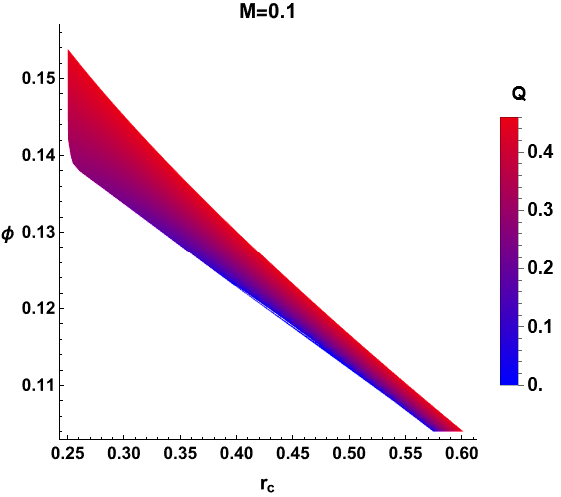}\\
  \end{tabbing}
\caption{ Left: The scalar field in terms of  the outer black hole radius for different values of the charge.  Right:  The scalar field in terms of the Cosmic horizon/Charged Nariai black hole for different values of the charge.}
\label{Fig3}
\end{center}
\end{figure}

  Exploiting some powerful CUDA techniques, mainly the dynamical parallelism and the  stream tails \cite{CUDA3,CUDA4},   we can determine the scalar field  as a function of  the horizon radius   by  considering  different values the black hole charges. Indeed, the numerical computations  are  illustrated in Fig.(\ref{Fig3}) by  considering  $M=0.1$.

This graphical  illustration   bridges  different parameters in the theory.   Globally,  it  shows that large horizons are less susceptible to changes in the charge. However,  such    illustrations seem to have opposite behaviors. For outer horizons $r_+$,  the scalar  field function is an increasing function. For certain  fixed event horizon values,  the scalar field increases by increasing the charge $Q$. Small outer horizons make this relation more apparent while larger ones allow smaller range of the scalar field. Regarding  the cosmic horizon,  the scalar field is  a decreasing function. Fixing the black hole  event horizon $r_c$,   it increases by increasing the charge $Q$. This behavior manifests clearly in the  the small horizon region.   In this way,   it should be interesting to  examine the scalar behavior in the  small horizon radius limit. It is denoted that such a limit indicates  the large scalar field values supporting   the presence  of the infinite towers of states.  Considering  such a  scalar field limit,   the computation   shows that  the cosmic horizon radius is found to be in the order of 
\begin{equation}
r_c\sim\sqrt{\alpha}.
\end{equation}
The small values of $\alpha$ suggest that the  small horizons probe UV scale where certain decompactification scenarios  could be  relevant. Similar arguments have been previously explored in some works including the ones reported in \cite{MS1,MS2}.

Motivated   by swampland  activities,   we would like to   establish a possible link between the scalar field and the black hole  charge.   This issue  will be considered in the fourcoming sections.
\section{WGC study of charged black holes}
\begin{figure}
\begin{center}
\includegraphics[width=9cm, height=6cm]{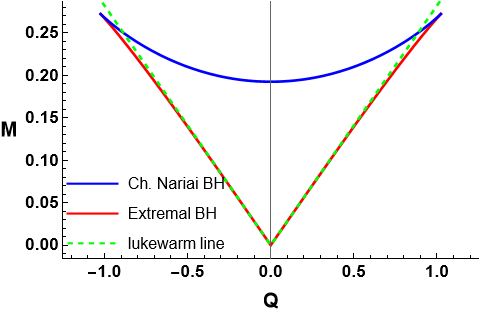}
\caption{Two branches of Eq.(\ref{Ext}) for $V(\phi) = 1.0$.}
\label{Fig11a}
\end{center}
\end{figure}

\begin{figure}
\begin{center}
\includegraphics[width=9cm, height=6cm]{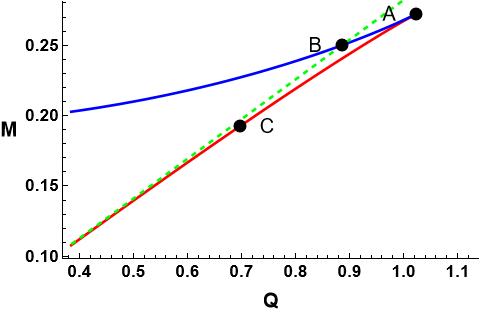}
\caption{The points $(Q,M)$  associated with  the stable sub-lukewarm heavy area for $V(\phi)=1.0$.}
\label{Fig11c}
\end{center}
\end{figure}

\begin{figure}
\begin{center}
\includegraphics[width=8cm, height=8cm]{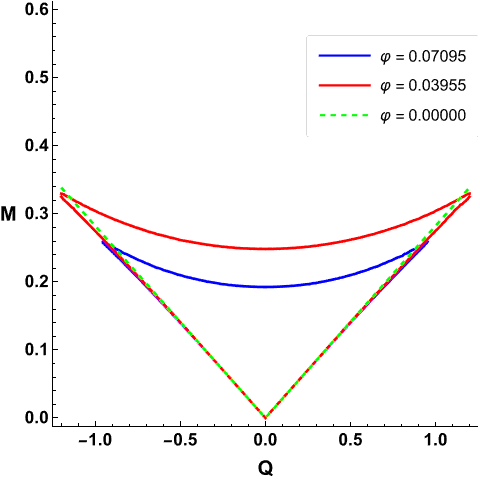}
\caption{The black hole mass $M$ in terms of  $Q$ for different values of the scalar field.}
\label{Fig11b}
\end{center}
\end{figure}

In this section, we would like to make contact with the swampland program. A special  emphasis is put  on  WGC. The latter is a good indicator of  QG  consistent EFTs. Besides, it prohibits the restoration of global gauge symmetries censoring  the naked singularities.  In other terms, one can state that an EFT is located in the landscape if it allows the  extremal black holes to decay into stable light states satisfying the ratio $\frac{Q}{M}>2\sqrt{\pi}$. Otherwise, it is said to be located in the swampland \cite{WGC1}.  To elaborate a link with WGC, the  relevant  object  is   the discriminant  of the quartic algebraic equation $g(r)=0$. This is found to be 
\begin{equation}
\Delta=4M^2-\frac{Q^2}{\pi}+36V(\phi)M^2\frac{Q^2}{\pi}-V(\phi)^2\frac{Q^6}{\pi^3}-2V(\phi)\frac{Q^4}{\pi}-108V(\phi)M^4.
\label{Ext}
\end{equation}
The locus $\Delta=0$  corresponds to the extremal curves in the black hole  parameter space $\lbrace Q,M\rbrace$.  Fig(\ref{Fig11a}) illustrates   the   extremal and   the charged Nariai black hole  curves.
The right half of this figure is known as the "shark fin" \cite{FL1,FL2}.  It involves  two different branches.  The   blue one, denoted by $M_{\mbox{\tiny Nariai}}(Q)$,  represents a  charged Nariai black hole with  a  horizon  given by $r_c$.   It is remarked that for  $r> r_c$,   the light cone of the observer frame will never reach.  The red  branch, denoted by $M_{\mbox{\tiny Extermal}}(Q)$, represents the extremal black hole solution where $r_{-}=r_{+}$. This means that  the inner black hole horizon coincides with the outer one.  For small $Q$ and  $M$ values,  this curve becomes indistinguishable from the lukewarm line represented by the dashed green curve.  It is safe to assume that the ratio $\frac{Q}{M}$ will always be grater than one.  This allows the existence  of a  certain  number of stable light states.\\
Lastly, the vertical line at $Q=0$ represents neutral black holes.
Combing  such  branches,  the black holes having  regular horizons and the  naked singularities can be  identified.  Precisely, the   black holes involving  regular horizons  are located inside $M_{\mbox{\tiny Nariai}}(Q)$  and   $M_{\mbox{\tiny Extermal}}(Q)$  curves.    However, the naked singularities are located in the outside.  For large values of  $Q$ and $M$,  it has been observed that these branches circumscribe an area where stable sub-lukewarm heavy states survive.     The geometrical configuration   is shown in Fig.(\ref{Fig11c}) for $V(\phi)=1.0$.
   It has been remarked  that this area is bounded from the top by ultra cold black hole solution (the intersection of the two branches) corresponding to the point $A\equiv(\sqrt{\frac{\pi}{3V(\phi)}}, \frac{\sqrt{2}}{3\sqrt{3V(\phi)}})$,  the intersection of the lukewarm and the charged Nariai curves at $B\equiv(\frac{1}{2}\sqrt{\frac{\pi}{V(\phi)}}, \frac{1}{4}\sqrt{\frac{1}{V(\phi)}})$ and the  bounded from the bottom by the point $C\equiv(\sqrt{\frac{2\pi-\pi\sqrt{3}}{\sqrt{3}V(\phi)}}, \frac{1}{\sqrt{27V(\phi)}})$.   This point  is obtained by assuming that the  sub-lukewarm heavy states correspond to the  black holes with masses equal or greater than the mass of  the neutral Nariai black holes. 
These heavy states are generally not desired, as the refined WGC specifically states that only light states should  be stable. Furthermore, heavy stable states contradict compactification arguments according to which their mass is directly proportional to the size of the compactified directions. Concretely, this issue reduces the impact of the inequality $\frac{Q}{M}>2\sqrt{\pi}$ on low energy physics. Effectively, one needs to limit the mass of stable charged particles.  To  overcome such an issue, the effect  of the scalar field might be needed.  Indeed, this  effect  helps to suppress the   sub-extremal stable heavy states. Explicitly the number of such states denoted by $N$ is proportional to the area of the (ABC) trigonal region. This number is  expressed as follows
\begin{equation}
N\sim\int^{\sqrt{\frac{\pi}{3V(\phi)}}}_{\frac{1}{4}\sqrt{\frac{\pi}{V(\phi)}}}(M_{\mbox{\tiny Nariai}}(Q)-M_{\mbox{\tiny Extermal}}(Q))dQ+\int^{\frac{1}{4}\sqrt{\frac{\pi}{V(\phi)}}}_{\sqrt{\frac{2\pi-\pi\sqrt{3}}{\sqrt{3}V(\phi)}}}(Q-M_{\mbox{\tiny Extermal}}(Q))dQ,
\end{equation}
which vanishes by shrinking    the distance moduli. Essentially,  one mediates the heavy stable states issue by assuming small field values. Similarly, the charged Nariai black hole curve is pushed upward along the mass axes by decreasing the value of the scalar field as shown  in Fig(\ref{Fig11b}). 
It has been remarked that the  large field values result  in small cosmic horizon values. This means that, for a   large moduli  distance, different reference frames exist outside each light cone, prohibiting the observation of infinite light state towers. In this context, it would be fair to assume some kind of relationships between the  moduli distance and the ratio between  the mass and  the charge.  In order to highlight such a correlation,  we provide  a thermodynamic study exploring thermodynamical  criticality behaviors in the  CUDA   assisted swampland context.


\section{ Bridging Swampland Program with Black Hole  Thermodynamics}
\begin{figure}
\begin{center}
\includegraphics[scale=0.7]{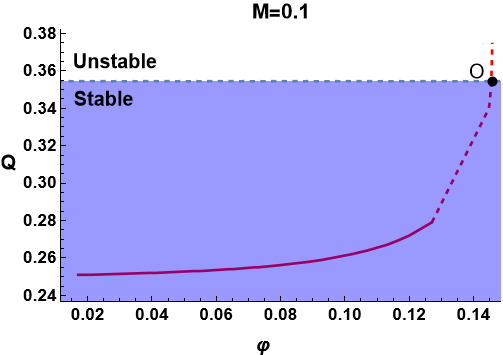}
\caption{Critical values of the scalar field in terms of the black hole charge for a  fixed  black hole mass being $M=0.1$.}
\label{Fig22}
\end{center}
\end{figure}

 In this section, we would like to establish  a link between the  swampland program and the black hole  thermodynamics using CUDA  numerical computations.   To do so, one needs to    determine some needed  thermodynamical  quantities.   Precisely,  we can obtain  the mass  in terms of  $ r_h$ by solving    the constraint  $g(r) = 0$ where   $r_h$  denotes either  the outer horizon  or  the  charged Nariai black hole horizon.   The computation provides 
\begin{equation}
M = \frac{1}{2}\left(r_h+\frac{Q^2}{4\pi r_h}-V(\phi)r^3_h\right).   
\end{equation}
Using  the usual calculations,   the  Hawking temperature with normalized Planck and Boltzmann constants $\hbar=k_B=1$  is found to be 
\begin{equation}
T =\frac{1}{4\pi r^3_h}\left(r^2_h-\frac{Q^2}{4\pi}-3V(\phi)r_h^4\right)\\ \label{xxx}.
\end{equation}
Taking in consideration that the specific volume to be $v=2r_h$, the state equation is expressed  then as follows
\begin{equation}
P=\frac{T}{ v }-\frac{1}{2\pi v^2}+\frac{2q^2}{\pi v^4},
\end{equation}
where one has  used  $q=\frac{Q}{\sqrt{4\pi}}$. The associated inflection points  should verify 
\begin{equation}
\frac{\partial P}{\partial v}=0, \hspace{6 pt} \frac{\partial^2P}{\partial v^2}=0.
\end{equation}
To obtain the critical values of the scalar field,  we  should solve such constraints.  Due to the   computation complexity,  intensive numerical  methods may be needed. This could be done  by taking advantage of the aforementioned CUDA techniques.  It  has been remarked that such an approach facilitates the determination of the  critical scalar field values, represented by the red curve for  the outer horizons of charged black holes and the  dashed red curve for horizons of the  charged Nariai black holes, as is illustrated in (\ref{Fig22}). 
It  has been  remarked that all points belonging to the red  curve  satisfy the critical ratio
\begin{equation}
\frac{P_cv_c}{T_c}=\frac{3}{8}
\end{equation}
showing that the  proposed black holes behave as    Van der Waals fluids\cite{17}. This correlation between the   scalar field and   the charge $Q$  is  essential to approach  the swampland program. In particular, the small values of the involved  parameters   designate  the stability regions separated by   the lukewarm limit. Consequently, unstable states correspond to values of the scalar field and  the charge where  one has  $\frac{Q}{M}>2\sqrt{\pi}$, which lie above the blue region.  In contrast, the scalar field   and the charge values  corresponding to stable states lie in the blue region, implying the inequality $\frac{Q}{M}<2\sqrt{\pi}$.    To distinguish the phases of large black holes from those of small black holes in the parameter space,  one should works  introduce  critical curves.  Relying on  WGC, these observations allow one  to approach  the stability of the aforementioned phases. Concretely, we observe that large black hole phases lay below the curve, while  the small ones are located above the curve. A careful examination reveals that the point $\mbox{O}\equiv(\phi_c\sim 0.143,Q_c\sim 0.355)$ has proven to be of great importance with regard to the WGC aspect. Elaborating further, for  the charge and the  field values greater than the values at $\mbox{O}$,  we observe that the phases of large black holes are separated into stable and unstable phases. However,   the light states are identified only with the ratio $\frac{Q}{M}>2\sqrt{\pi}$. On the contrary, for values smaller than the values associated  with  the point $\mbox{O}$ the only possible large black hole phases are the stable ones.  On the other hand, the small black hole phases can satisfy the both cases of the WGC being  not prohibited for light stable states.\\

It has been suggested that the small black hole phases appear to favor the inequality $\frac{Q}{M}>2\sqrt{\pi}$, associated with light states. This seems to be consistent with certain findings \cite{WGC1}, where  the small black holes behave as stable light particles satisfying the WGC. Regarding  the heavy states, we expect that for scalar field values greater than the value of the stabilization point $\mbox{O}$, large black holes of mass $M=0.1$ can be endowed with more charge leading to instability behaviors. Consequently, the only possible decay channel is light states adhering to the WGC. However, the  small values of the moduli support only large black holes satisfying the  extremal inequality $M>\frac{Q}{2\sqrt{\pi}}$.

In connection with Fig(\ref{Fig3}), the aforementioned observations explain the interplay between the moduli distance and the black hole charge in a clear manner. Explicitly, the moduli distance could indicate at what stage of the observable universe charged black holes stabilize.  

\section{ Conclusions}

In this work, we have  investigated  the  swampland program  in the context of  the stringy black hole  thermodynamics with help of CUDA numerical computations.  First, we have reconsidered  the study   of charged black holes submerged in an inflationary potential, endowed with a scalar field that couples to  a stringy correction via the Maxwell-Einstein-Hilbert action  involving GB contributions.  Using a numerical  approach, we  have examined  the physically relevant roots of the black hole  metric function $g(r)$. To  examine the effects of the different parameters (charge, moduli distance, mass...) on the physical quantities, we have produced robust analytic computations achieved by capitalizing on CUDA enabled parallel computing. As a result, the swampland program has been motivated in order to unveil more information regarding such  parameters mainly the relation between the black hole charge, the mass and the moduli distance.  Precisely, we have studied extremal limits of such black holes.  The extremal curve is only one solution of the vanishing discriminant $\Delta = 0$ while the other solution is represented by the charged Nariai black hole curve. Exposing the relation between the black hole charge $Q$ and  the mass,  we have approached  the values of  the mass and  the charge corresponding to black holes with regular horizons. In particular,  we have observed that  these  obtained curves allow the existence of a number of undesirable sub-lukewarm stable heavy states.  In particular, we have facilitated this process by taking into account the distance of moduli.  Explicitly,  we have revealed that  the number of these states vanishes by shrinking the moduli distance. Moreover,  we have remarked  that the cosmic horizon  can be  sent upwards in the mass direction, while the extremal curve   becomes less  distinguishable from the lukewarm curve. Additionally, we have   clearly elaborated  a bridging scenario  between the black hole  charge and the moduli distance.  Using the developed CUDA techniques,  we have  related the black hole charge to the moduli distance by exploiting  the black hole thermodynamics. Accordingly, we  have extracted  the critical values of the moduli distance in terms of the electric  charge  distinguishing   between small and large black hole phases. Supported by the WGC, we have assigned each phase a stable or unstable character according to their  corresponding  charge  and  the  moduli  distance. Concretely, we have  identified the point at which large black holes stabilize. Furthermore, for a moduli distance and charge smaller than  the value at the stabilization, we have observed that  all large black hole solutions   satisfy   extremal bounds. However, the  WGC instability starts to manifest for  relevant charge and moduli distance contributions.

This work leaves certain open questions.    We expect that the present    approach  could be  adaptable to a broad variety of  charged black holes including the regular ones. The extension  to such solutions    could be  issues for  future works.  We anticipate
that   the CUDA techniques could play a key role in unveiling more aspects of the swampland program. This will be addressed elsewhere.
\section*{Acknowledgements}
The   authors  SEB and HB  would like to dedicate the present  work to  the memory of   Fatima Addoud,  AB  mother.

\end{document}